\documentclass{appolb}
\usepackage{epsfig}
\usepackage{amssymb}

\newcommand{\be}{\begin{equation}}
\newcommand{\ee}{\end{equation}}
\newcommand{\ba}{\begin{eqnarray}}
\newcommand{\ea}{\end{eqnarray}}
\newcommand{\fr}[2]{{\frac{#1}{#2}\,}}

\newcommand{\rmi}[1]{{\mbox{\scriptsize #1}}}

\begin{document}
\title{The QCD Equation of State --- From Nuclear Physics to Perturbation Theory%
\thanks{Presented at the workshop \textit{Excited QCD 2011}, Lec Houches, Feb.~20-25, 2011.}%
}
\author{Aleksi Vuorinen
\address{Faculty of Physics, University of Bielefeld, D-33501 Bielefeld, Germany}
}
\maketitle
\begin{abstract}
In this talk, we briefly review the current understanding of the behavior of the QCD equation of state throughout the phase diagram. Special emphasis is given to regions of phenomenological interest, and a number of important open questions as well as directions of ongoing research are pointed out. These include in particular the region of low temperatures and (moderately) high densities, where at the moment we have extremely few first principles tools available.
\end{abstract}

\section{Introduction}

Apart from being of obvious theoretical interest, the behavior of the Equation of State (EoS) of QCD as a function of temperature $T$ and baryon chemical potential $\mu_B$ is of considerable phenomenological importance in a variety of physical systems. Through Einstein's equations, the high-temperature and low-density EoS governed the expansion and cooling rates of the universe when its age was of the order of a few tens of microseconds, and through the expansion rate it thus had an effect on \textit{e.g.~}the decoupling times of various dark matter candidates. In heavy ion physics, the success of the hydrodynamic description of the collision relies on the availability of accurate information on the energy density and pressure of the underlying theory. And finally, the properties of compact stars --- in fact, their entire composition --- are extremely sensitive to the EoS of cold and dense nuclear and quark matter.

In this talk, we will review what is currently known about the EoS of QCD in various parts of the $\mu_B$-$T$ phase diagram, covering both the confined and deconfined regimes and extending from zero density all the way to zero temperature. Conceptually, the setup of the problem is extremely simple: One takes the Euclidean functional integral corresponding to the partition function (or grand potential) of the theory,
\ba
\Omega(T,\{\mu_f\},\{m_f\})&=& -T\log \int \mathcal{D}\bar{\psi}\mathcal{D}\psi\mathcal{D}A_\mu e^{-\int_0^\beta d\tau \int d^3 x \mathcal{L}_\rmi{QCD}}, \label{omega}\\
\mathcal{L}_{\rmi{QCD}}&=&\frac{1}{4}F^a_{\mu\nu}F^a_{\mu\nu}+\bar{\psi}_f(\gamma_\mu D_\mu+m_f
-\mu_f \gamma_0)\psi_f, \label{LQCD}
\ea
evaluates it one way or another, and subsequently uses standard relations to obtain predictions for a number of equilibrium thermodynamical quantities ranging from the pressure to the quark number, entropy and energy densities
\ba
pV&=&-\Omega,\\
n_f V &=& -\partial_{\mu_f}\Omega,\\
sV&=&-\partial_T\Omega,\\
\varepsilon &=& -p+Ts + \mu_f n_f.
\ea
The real challenge of course becomes, how to choose the optimal method for doing this in different parts of the phase diagram, and this question is what we will be mostly concerned with. There is no one single method to cover the entire $\mu_B$-$T$ plane, but as we will argue below, combining results from various existing techniques provides us with a quantitative handle on the behavior of the EoS \textit{almost} everywhere.

To make the following discussion more transparent and physically intuitive, we choose to parametrize the temperature and baryon chemical potential by a radial and angular variable. These are defined by
\ba
\rho &\equiv& \sqrt{T^2+\fr{\mu_B^2}{6\pi^2}},\quad \theta\;\equiv\; {\rm arctan}\fr{\mu_B}{\sqrt{6}\pi T},\\
T&=&\rho\cos\theta,\quad \mu_B\;=\;\sqrt{6}\pi \rho \sin\theta,
\ea
where $\rho$ is roughly proportional to the fourth root of the energy density of the free theory, while the angle $\theta$ measures the deviation from the temperature (zero density) axis.

The roles of the variables are clear: Due to the running of the $\alpha_s$, the energy scale $\rho$ measures how strongly coupled the system is. Roughly speaking, at $\rho\lesssim 100$ MeV we are deep in the confined phase of the theory, at 100 MeV $\lesssim \rho\lesssim$ 1 GeV the system resides in an intermediate region characterized by a deconfinement phase transition (or crossover) and complicated dynamics, and at $\rho\gtrsim$ 1 GeV, we enter a regime where a description in terms of weakly interacting, deconfined quasiparticles becomes feasible. The angle $\theta$ on the other hand separates two phenomenologically interesting regions: The values $\theta\lesssim 1$ are relevant for the physics of the early universe and heavy ion collisions, while in the matter found inside neutron stars, we have to a very good accuracy $\theta = \pi/2$. Thus, while it may be theoretically interesting to consider the EoS for other values of $\theta$ as well, most of our attention will in what follows be divided between these two cases.

\section{Small $\theta$: From Hadron Gas to Quark-Gluon Plasma}

For baryon chemical potentials of at most the order of the temperature, the two limits of low and high energy densities, respectively $\rho\lesssim 100$ MeV and $\rho\gtrsim 1$ GeV, are conceptually relatively simple. At low energies, we are dealing with a dilute gas of hadrons that may be described in terms of hadron resonance gas (HRG) models, while at high $\rho$, the system comprises a (relatively) weakly coupled quark-gluon plasma that is amenable to treatment via resummed perturbation theory. In the regime between these two limits, one may attempt to gain qualitative understanding of the dynamics of the system via various types of effective theories and models, but the ultimate tool for quantitative information is clearly lattice QCD.

The success of the hadron resonance gas models, originally proposed in Ref.~\cite{Hagedorn:1965st}, is based on confinement. At sufficiently low energy density, the system comprises a dilute gas of hadrons, in which the strong interactions are very efficiently confined inside these particles, making the system in practice a free Stefan-Boltzmann gas. Raising the value of $\rho$, and thus introducing more degrees of freedom, one must begin to take into account both the interactions between the hadrons and the fraction of spatial volume occupied by them, usually according to the Van der Waals prescription. Eventually, these effects lead to sizable uncertainties in the results. Nevertheless, current calculations, which typically include ${\mathcal O}(100)$ hadrons up to masses of 2-3 GeV, have been extended to close to the phase transition region, and thus the results now have a finite overlap region with lattice simulations. For details of these calculations as well as up-to-date results, see \textit{e.g.~}Ref.~\cite{Huovinen:2009yb} and references therein.

For $\rho$ between roughly 100 MeV and several times the critical temperature $T_c$ of the deconfinement transition, the method of choice for quantitatively reliable results is lattice QCD. For a long time, the main challenge in lattice simulations was to reach physical quark masses and a wide enough temperature range, but recent years have witnessed enormous progress in both of these fronts \cite{Cheng:2009zi,Borsanyi:2010cj}. As demonstrated \textit{e.g.~}in Fig.~2 of Ref.~\cite{Borsanyi:2010kj}, there is now an impressive agreement between current lattice data and the predictions of hadron resonance gas models at temperatures slightly below $T_c$. In addition, the discrepancies between the predictions of the different lattice groups regarding the EoS and the critical temperatures of the chiral and deconfinement transitions have recently significantly decreased. At the moment, the main limitation of the lattice approach to QCD thermodynamics is clearly the sign problem, prohibiting simulations at baryon chemical potentials much larger than the temperature. For recent progress in this direction, see \textit{e.g.~}Ref.~\cite{Karsch:2010hm} and references therein.

Even higher temperatures and energy densities, $\rho\gtrsim 1$ GeV, witness lattice simulations becoming more and more expensive\footnote{See, however, Ref.~\cite{Endrodi:2007tq} for lattice results extended to high temperatures.} while perturbation theory starts to work better and better. While the pure weak coupling expansion of the QCD pressure, at the moment worked out up to the partial $g^6$ term \cite{Kajantie:2002wa,Kajantie:2003ax,DiRenzo:2006nh,Gynther:2009qf}, shows relatively poor convergence, the situation can be greatly improved using resummed perturbation theory. The most natural platform for this is offered by dimensionally reduced effective theories, in addition to which very promising results have recently been obtained using Hard Thermal Loop resummed perturbation theory \cite{Blaizot:2003iq,Andersen:2011sf}. The perturbative approach has in addition the virtue of being easily generalizable to finite chemical potentials \cite{Vuorinen:2003fs}, as even dimensional reduction has been shown to lead to quantitatively reliable results as long as $\mu_B\lesssim T/g$ \cite{Ipp:2006ij}.

\section{$\theta\approx \pi/2$: Cold Nuclear/Quark Matter}

The region of the QCD phase diagram characterized by finite baryon density and (negligibly) small temperatures is in many ways more problematic than that of small $\theta$. For one thing, lattice QCD is no longer available as a computational method, and one must thus base one's approach on a combination of weak coupling techniques and model calculations. In addition, it appears that the exact phase structure of the theory has a highly non-trivial form at small temperatures, and it is only at asymptotically large chemical potentials that the true nature of the deconfined phase is known. There, the pairing instability of the perturbative ground state drives quarks on the Fermi surface to pair in a way that ties their color and flavor indices together, resulting in the so-called Color Flavor Locked (CFL) superconductor. The questions of how far down the $\mu_B$ axis this phase extends to and which of the several proposed candidate phases ultimately replaces it, are, however, still open. Luckily, not all equilibrium thermodynamic quantities, in particular the EoS, are extremely sensitive to these details.

Starting again from small energy densities and the nuclear matter phase, we enter a corner of the phase diagram that is under relatively good quantitative control. This is not due to first principle calculations, however, but rather to the abundance of experimental data on low energy nucleon-nucleon scattering, enabling an accurate modeling of the dilute nuclear matter EoS. In close analogy to the HRG approach, the quantitative accuracy of these models begins to suffer as one increases $\rho$ and approaches the deconfinement transition region. In the case of cold and dense nuclear matter, the primary uncertainties in the EoS are due to the unknown composition of the matter (the conjectured presence of \textit{e.g.~}hyperons or kaon condensation), and minor ones to the details of the model calculations such as the exact form of the respective variational ansatz or the effect of neglecting the simultaneous interactions of more than two nuclei. For further details of these computations as well as for references regarding the nuclear matter EoS, the reader is asked to consult Refs.~\cite{Akmal:1998cf,Schulze:2006vw,Glendenning:1998zx}.

In the case of asymptotically large energy densities, $\rho\gg 1$ GeV, the physical ground state of QCD is a CFL superconductor. It is characterized by a (perturbatively) exponentially small energy gap $\Delta\sim e^{-\#/g}$, and it is thus natural that the effects of the gap can be formally neglected in a weak coupling expansion of the EoS. In any practical applications, \textit{e.g.~}when discussing compact stars, this is, however, certainly not good approximation, and one should in addition pay close attention to quark mass threshold crossings. Both of these effects have been included in the recent three-loop perturbative computation of Ref.~\cite{Kurkela:2009gj}. The results of this calculation demonstrate that while the zero-temperature pressure does exhibit somewhat better convergence properties than the high-$T$ one, the uncertainties in the results become sizable when one approaches densities that might be realized in nature.

It should not come as a surprise that in the $\theta\approx \pi/2$ case, the most problematic region is by far that of intermediate densities, 100 MeV $\lesssim \rho\lesssim$ 1 GeV. In this regime, neither the nuclear model approach nor perturbation theory is applicable, but obtaining accurate information on the behavior of the EoS is nevertheless crucial in order to understand the structure of compact stars. Two very different lines of attack have been suggested to overcome these difficulties. One may either trust one of the several models that have been built to describe the relevant physics in this regime (usually following the guidance of symmetry principles) \cite{Buballa:2003qv}, or interpolate between the trusted EoSs of dilute nuclear and dense quark matter \cite{Kurkela:2009gj}. Recent advances in compact star observations have in addition made it possible to directly rule out sizable regions of their mass-radius plane. One day, this may even lead to the solution of the inverse problem, \textit{i.e.~}predicting the EoS of cold nuclear and quark matter solely from experimental data (see \textit{e.g.~}Refs.~\cite{Steiner:2010fz,Kurkela:2010yk} for a discussion of this).

\section*{Acknowledgements}

A.V. was supported by the Sofja Kovalevskaja program of the Alexander von Humboldt foundation.

\end{document}